\documentclass[conference]{IEEEtran}
\usepackage[top=0.7in, bottom=0.9in, left=0.6in, right=0.6in]{geometry} 
\usepackage{amsmath}
\usepackage{graphicx}
\usepackage{cite}
\usepackage{url}
\usepackage{color}
\usepackage{subfig}
\usepackage{caption}

\usepackage{graphicx,color}
\usepackage{textcomp}
\usepackage{xcolor}
\usepackage{array}
\usepackage{multirow}
\usepackage{makecell}
\usepackage{float}
\usepackage{adjustbox}
\usepackage{booktabs}
\usepackage{longtable}
\usepackage{tabularx}
\usepackage{afterpage}
\usepackage{stfloats}

 \usepackage{fancyhdr}
 \fancypagestyle{firststyle}{\fancyhf{}
	 		\fancyhead[L]{T. S. Rappaport, D. Shakya, and M. Ying, ``Point Data for Site-Specific Mid-band Radio Propagation Channel Statistics in the Indoor Hotspot (InH) Environment for 3GPP and Next Generation Alliance (NGA) Channel Modeling",\textit{IEEE International Conference on Communications (ICC), }Montreal, Canada, Jun. 2025, pp. 1--6.}
	 	}
 \usepackage{times}

\begin{document}
	
	\title{Point-Data for Site-Specific Mid-band Radio Propagation Channel Statistics in the Indoor Hotspot (InH) Environment for 3GPP and Next Generation Alliance (NGA) Channel Modeling}
	
	\author{
		\IEEEauthorblockN{Theodore S. Rappaport, Dipankar Shakya, and Mingjun Ying}
		\IEEEauthorblockA{NYU WIRELESS, Tandon School of Engineering, New York University, Brooklyn, NY, 11201\\
			\{tsr, dshakya, yingmingjun\}@nyu.edu}
		\thanks{This research is supported by the NYU WIRELESS Industrial Affiliates Program and the NYU Tandon School of Engineering graduate fellowship.}
		\vspace{-15pt}
	}
	
	\maketitle
	
	\thispagestyle{firststyle}
	
	\begin{abstract}
		Extensive work has been carried out in the past year by various organizations in an effort to determine standardized statistical channel impulse response (CIR) parameters for the newly-released \textcolor{black}{FR3} mid-band spectrum (7.25 GHz -- 24.25 GHz) \cite{Shakya2024gc, Shakya2025wcnc, Shakya2024ojcoms, Kang2024OJCOM, Huawei2021}. In this work, we show that the wireless community currently lacks a unified method for presenting key parameters required for transparency and utilization by several constituencies when presenting propagation data for use by standard bodies or third parties to create statistical CIR models. This paper aims to solve the existing problem by offering a standard method to provide key propagation parameters in a point-data format that supports both statistical and site-specific channel characterization. The proposed method offers tremendous promise when data contributors use the minimum agreed-upon measurement and processing specifications such as bandwidth, antenna beamwidth, and noise threshold level. As shown here, the point-data format enables multiple contributors to create channel model standards or pool measurement data to create larger datasets for exploring ray-tracing (e.g. site-specific) channel modeling or training in AI/ML propagation work, and to ensure the most accurate model using a larger dataset that is continually expanded through measurement contributions. The point-data approach includes site-specific point-by-point propagation data while readily supporting the creation of commonly-used cumulative distribution function (CDF) plot. The indoor hotspot (InH) datasets collected in Spring 2024 at 6.75 GHz and 16.95 GHZ by NYU WIRELESS \cite{Shakya2024gc, Shakya2025wcnc, Shakya2024ojcoms} are provided for the first time in point-data form, to augment statistical models previously presented solely as CDFs, in order to demonstrate how a standardized approach to measurement data could allow others to utilize the site-specific locations and key channel parameters observed at each location, to better understand, vet, and build upon statistical or site-specific CIRs from the contributions of many different data sources.
	\end{abstract}
	
	
	\section{Introduction}
	
	The global standards bodies, including 3GPP and the Next Generation Alliance (NGA) in North America, have initiated channel modeling activities to harmonize the existing statistical radio propagation channel models for both Frequency Range 1 (FR1) at microwave frequencies (410 to 7125 MHz) \cite{3GPPTR38901} and Frequency Range 2 (FR2) bands at millimeter-wave frequencies (24.25 GHz to 52.6 GHz) \cite{3GPPTR38900}. 
	
	Within the past year, regulatory agencies throughout the world have made the mid-band frequencies between 7.25 GHz and 24.25 GHz—denoted as frequency range 3 (FR3)—available for use by the global cellular industry \cite{Kang2024OJCOM, Huawei2021, NTIA2024, 3GPP3820}. This new allocation and the promise of cellular service in these spectrum allocations have set off a flurry of radio propagation and channel measurement activities worldwide. Recent research efforts aimed at characterizing and modeling the new mid-band spectrum can be found in \cite{Shakya2024gc, Shakya2024gc2, Shakya2024ojcoms, Shakya2025wcnc, zhou2017iet, Oyie2018ia, janssen1996tc, Kim2014tap,al2018wcmc,diakhate:2017:millimeter-wave-outdoor-to-indoor}.
	
	Standards bodies such as 3GPP Radio Access Network (RAN) Technical Specification Group (TSG) and NGA have monthly meetings dedicated to experimental and theoretical findings of mid-band spectrum propagation characteristics. These findings usually provide a limited number of specific observations \cite{r12403280, r12402407,rappaport2017investigation} or statistical models of particular propagation parameters (e.g., path loss, time delay spread, etc.) over a wide range of measured locations \cite{Shakya2024gc, Shakya2024ojcoms,rappaport2017investigation}. However, only providing the statistical models makes it difficult for industry and standard bodies to effectively use them to create standards or compare parameters using the entirety of data. In this work, we provide a standard method for providing propagation parameters on a location-by-location, point-by-point basis. The concept of providing point-data was first described in \cite{rappaport2011open}, and was developed in an early stage for mmWave channel models to present path loss \cite{maccartney2015indoor,maccartney2015outdoor,rappaport2015tcom}, however, it was never adopted to any large extent by standard bodies.
	
	\section{Prior Work at Mid-band FR3}
	{\color{black}
		Several studies have reported empirical indoor path loss and delay spread measurements in the FR1(C) and FR3 frequency bands \cite{Shakya2024ojcoms, Sun2016tvt, zhou2017iet, Wei2024vtc, Oyie2018ia, Kim2014tap, Yin2017ia, Shakya2024gc, janssen1996tc, Rappaport1992tc, Miao2023jsac}. Authors in \cite{Shakya2024ojcoms} provide a summary of previous measurements in Table 2, while providing details about an extensive channel propagation measurement campaign conducted at the NYU WIRELESS Research Center using a 1 GHz bandwidth sliding correlation channel sounder at 6.75 GHz and 16.95 GHz. Using a close-in (CI) free space path loss (FSPL) model\cite{Sun2016tvt} with 1 m reference distance, the measurement campaign revealed an omnidirectional (omni) path loss exponent (PLE) of 1.34 in line-of-sight (LOS) and 2.72 in non-LOS (NLOS) at 6.75 GHz, and 1.32 in LOS and 3.05 in NLOS at 16.95 GHz. Zhou \textit{et al.} \cite{zhou2017iet} conducted propagation measurements in an office corridor at 11 and 14 GHz using a vector network analyzer (VNA)-based channel sounding system with biconical antennas. Their measurements found that the PLE values using the CI FSPL model with 1 m reference were 1.52 and 1.59 in LOS conditions, and 3.06 and 2.76 in NLOS conditions at 11 and 14 GHz, respectively. The RMS delay spreads (DS) were 19.5 ns and 17.9 ns in LOS at 11 GHz and 14 GHz, respectively, while they measured 23.43 ns and 22.03 ns at 11 GHz and 14 GHz in NLOS. Authors in \cite{Shakya2024ojcoms} observed omni RMS DS of 37.7 ns and 48 ns at 6.75 GHz and 22.1 ns and 40.7 ns at 16.95 GHz, using a threshold of the greater between 25 dB below the peak or 5 dB above noise floor, revealing a decreasing DS with increasing frequency.
		In a separate study, Wei \textit{et al.} \cite{Wei2024vtc} conducted measurements in university corridors at 6 GHz using a MIMO array with a 100 MHz bandwidth and a sliding correlation channel sounder. The RMS DS observed were around 21 ns in LOS and approximately 39 ns in NLOS (unreported threshold). Oyie \textit{et al.} \cite{Oyie2018ia} used continuous wave measurements in a university hallway at 14 and 22 GHz with horn antennas (19.5 and 22 dBi gain, respectively) and recorded PLEs of 1.6 and 1.7 in LOS, based on a CI FSPL model with 1 m reference distance. Kim \textit{et al.} \cite{Kim2014tap}, using a MIMO channel sounder transmitting unmodulated tones over a 400 MHz bandwidth at 11 GHz reported a PLE of 1.18 in LOS and 3.28 in NLOS from a CI FSPL model fit on the path loss with a 1 m reference distance. Swept narrowband measurements from 14-17 GHz using a VNA in \cite{Yin2017ia} observed RMS DS of 2.7 ns and 12.5 ns in LOS and NLOS, respectively. The same measurements in \cite{Yin2017ia} revealed RMS Angular Spread (AS) of 5.4$^\circ$ for the directional measurements with a 10$^\circ$ half-power beamwidth antenna. The campaign in \cite{Shakya2024gc} observed wide omni RX AS of 40.9$^\circ$ LOS and 58.2$^\circ$ NLOS at 6.75 GHz and 34.2$^\circ$ LOS and 43.5$^\circ$ NLOS at 16.95 GHz, suggesting a spatial richness of multipath. 
		
		VNA-based propagation measurements were reported in \cite{janssen1996tc} in office and laboratory environments at frequencies of 2.4, 4.75, and 11.5 GHz with bandwidths of 500 MHz and 1 GHz. They reported LOS PLEs of 1.86, 1.98, and 1.94, and NLOS PLEs of 3.33, 3.75, and 4.46 using the CI FSPL mod 1 m reference distance for the respective frequencies. Authors in \cite{Rappaport1992tc} observed a PLE of approximately 1.8 for indoor environments using CI FSPL model (1 m ref distance) in LOS at 1.4 GHz and 4 GHz, suggesting a waveguiding effect at these frequencies. 
		In another study, Miao \textit{et al.} \cite{Miao2023jsac} analyzed RMS delay spreads in outdoor-to-indoor (O2I) environments at 3.3, 6.5, 15, and 28 GHz, noting larger delay spreads for O2I than outdoor scenarios, but with no clear frequency trend.
		
	}

	\section{A New Approach for Propagation Datasets}
	
	None of the papers reported in the literature have offered a standardized approach that fosters understanding for all key parameters using the site-specific nature of the reported measurements. \textcolor{black}{The point-data concept was first described in \cite{rappaport2011open}, and while there were early adoptions of the point-data format for location-wise PL data in \cite{maccartney2015outdoor, rappaport2015tcom, maccartney2015indoor} for 28, 38, and 73 GHz, the approach was never widely adopted for standardization.} Sadly, 3GPP and NGA standard bodies do not allow large constituencies to understand particular measurement values of specific parameters at individual points or locations in a well-defined site-specific environment. Rather, statistical distributions, such as CDFs or scatter plots, are generally offered, making it challenging to incorporate a contributor's measurement data into a larger pool of data to improve or learn channel characteristics, or to rapidly match particular measured values to specific locations or points on a map in order to improve site-specific channel modeling and prediction.
	
	
	\textcolor{black}{At NYU WIRELESS, an intense measurement campaign, using a 1 GHz wideband channel sounder at 6.75 GHz and 16.95 GHz was undertaken from February 20 to May 30, 2024, in and around the NYU Brooklyn campus. In total, over 240 GB of propagation data was collected from the measurements in indoor, outdoor, and factory environments. A 15 dBi gain, 30$^\circ$ half power beam width (HPBW) antenna, and a 20 dBi gain and 15 $^\circ$ HPBW antenna were used for 6.75 GHz and 16.95 GHz measurements, respectively, where omnidirectional channel measurements were carried out via meticulous HPBW-stepped antenna rotations~\cite{Shakya2024gc, Shakya2024gc2, Shakya2024ojcoms, Shakya2025wcnc, Ying2025tcom, Ying2025icc,Shakya2022icc}.}

\begin{figure*}[htbp] 
		\centering
		\subfloat[Indoor InH Map--NYU WIRELESS Research Center, Brooklyn, NY, USA]{%
			\includegraphics[width=0.67\textwidth]{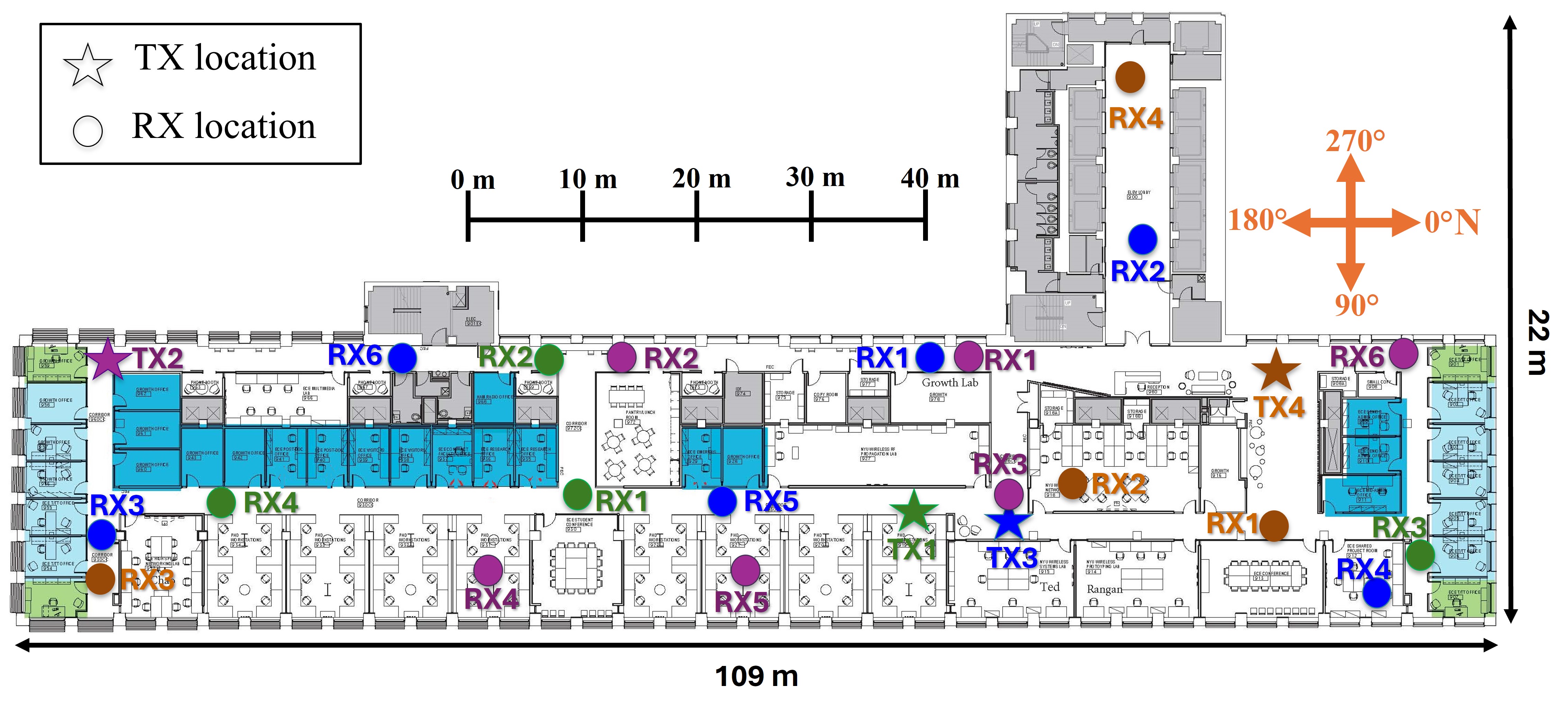} 
			\label{fig:inh_map}
		}\\
		\subfloat[Factory InF Map--NYU MakerSpace, Brooklyn, NY, USA]{%
			\includegraphics[width=0.45\textwidth]{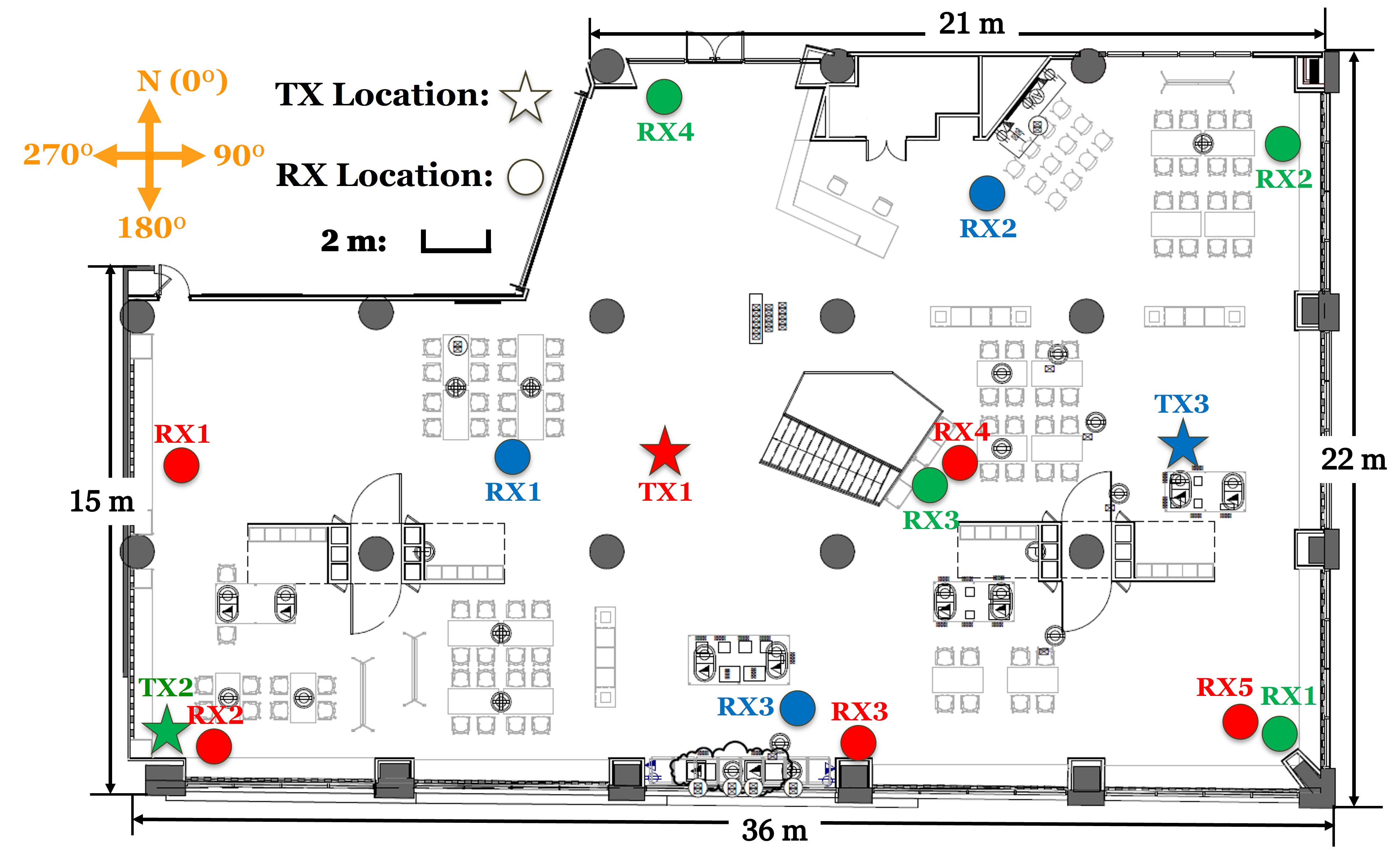} 
			\label{fig:inf_map}
		}
		\subfloat[Outdoor UMi Map (up to 200 m)--MetroTech Commons, Brooklyn, NY, USA]{%
			\includegraphics[width=0.35\textwidth]{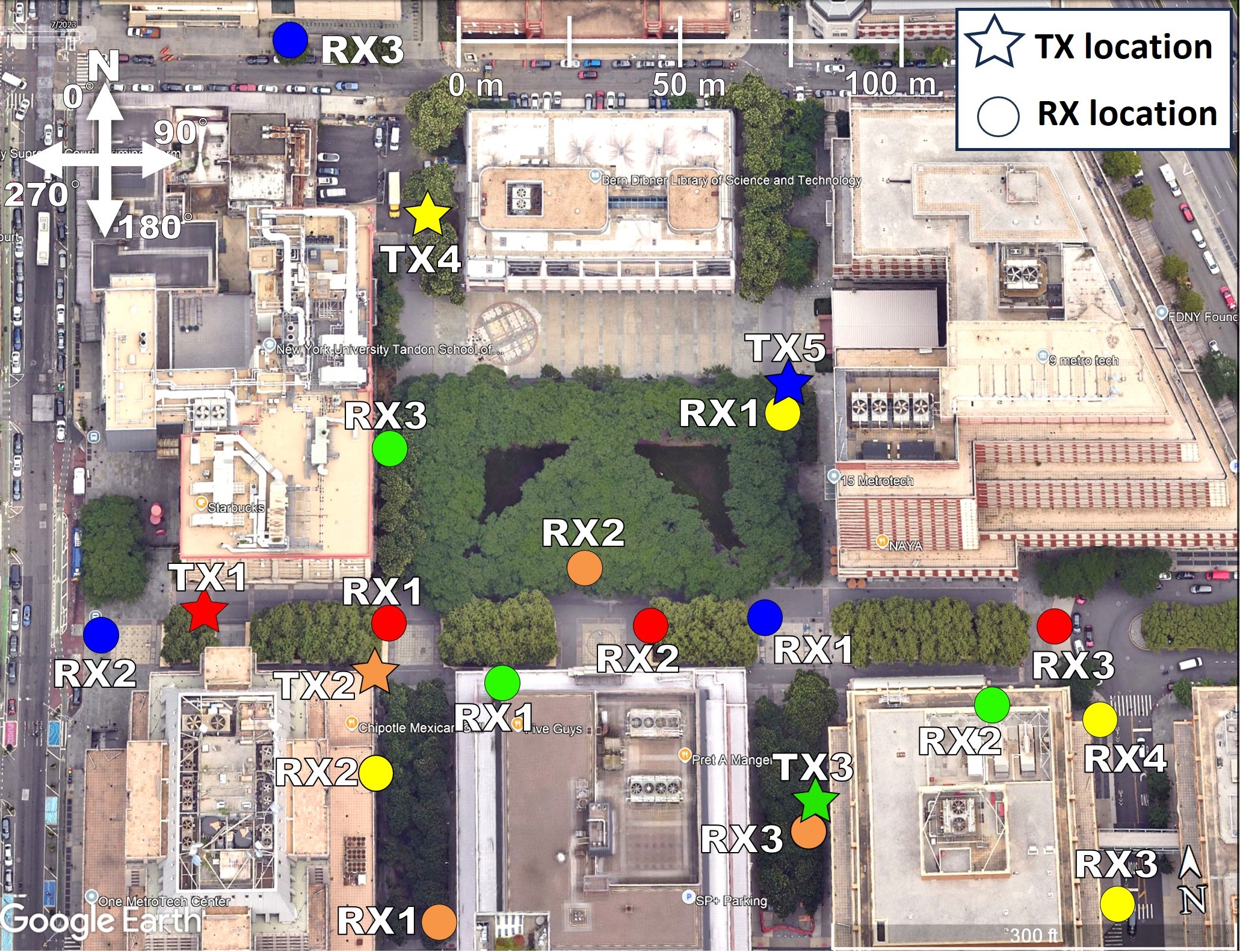} 
			\label{fig:umi_map1}
		}\\
		\subfloat[Outdoor UMi Map 2 (up to 1 km)--MetroTech, Brooklyn, NY, USA]{%
			\includegraphics[width=0.7\textwidth]{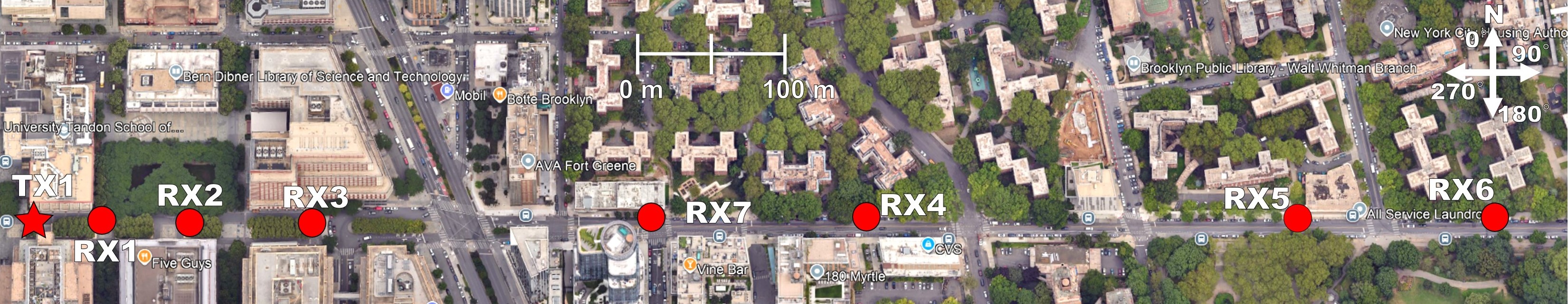} 
			\label{fig:umi_map2}
		}  
		\caption{Maps of the environment for the NYU FR1(C) and FR3 channel propagation measurement campaigns: (a) Indoor Hotspot (InH), (b) Indoor Factory (InF), (c) Urban Microcell (UMi) (up to 200 m), and (d) UMi (up to 1 km).}
		\label{fig:measurement_maps}
		\vspace{-15 pt}
	\end{figure*}

	\textbf{Indoor Hotspot (InH)} environment in Fig. \ref{fig:measurement_maps}(a) shows the NYU WIRELESS Research Center, 370 Jay Street, Brooklyn, NY. This InH environment features a typical office layout with multiple office rooms and corridors. Total 20 TX-RX pairs were measured, as indicated as points on the map of Fig. \ref{fig:measurement_maps} (a), with T-R separations ranging from 13 m to 97 m. The measurements included seven LOS and 13 NLOS scenarios, demonstrating varied indoor conditions across different office rooms and hallways. 110 GB of propagation data were collected in this environment~\cite {Shakya2024gc, Shakya2024gc2, Shakya2024ojcoms,Shakya2025wcnc,Ying2025tcom}.
	
	\textbf{Indoor Factory (InF)} environment shown in Fig. \ref{fig:measurement_maps}(b) illustrates the NYU MakerSpace located at 6 MetroTech, Brooklyn, NY. This InF environment has an open layout resembling a manufacturing floor equipped with 3D printers, laser cutters, CNC milling machines, imaging tools, PCB construction stations, heating tools, etc. The campaign included 12 TX-RX pairs, with distances ranging from 9 m to 38 m, capturing five LOS and seven NLOS scenarios with no outages. 55 GB of propagation data were collected in this environment~\cite{Ying2025icc, Ying2025tcom}.
	
	\textbf{Urban Microcell (UMi)} environment captured in Fig. \ref{fig:measurement_maps}(c) and Fig. \ref{fig:measurement_maps}(d) is an outdoor area around the MetroTech commons in Brooklyn, which is representative of the dense urban environment of New York City. A total of 20 TX-RX pairs were measured, with separation distances ranging from 35 m up to 1 km, including six LOS and 14 NLOS scenarios, as indicated specifically by points on the map shown in Fig. \ref{fig:measurement_maps} (c) and Fig. \ref{fig:measurement_maps} (d). The two maps shown in Fig. \ref{fig:measurement_maps} (c) and (d) depict all the TX-RX separations:  Fig. \ref{fig:measurement_maps} (c) shows TX-RX locations separated up to 216 m covering the Tandon School of Engineering courtyard, and Fig. \ref{fig:measurement_maps} (d) with TX-RX distances extending up to 1 km. Two outages (TX1-RX6 and TX4-RX3) are observed during the UMi measurement campaign. 75 GB of propagation data were collected in this environment~\cite{Shakya2025icc, Ying2025tcom}.

	The extensive InH statistical results (e.g. CDFs and scatter plots) of the propagation parameters observed within the shown in Fig. \ref{fig:measurement_maps} (a) were recently published in \cite{Shakya2024gc, Shakya2024gc2, Shakya2024ojcoms, Shakya2025wcnc}, but when offering these results to industrial participants of 3GPP and NGA, it became clear that companies found it difficult, if not impossible, to use the statistical results of the measured data given in \cite{Shakya2024gc, Shakya2024gc2, Shakya2024ojcoms} to compare with their own measurements, or to build a pool of statistical parameters from many different contributors from which to derive a statistical model from a global set of measured data from many contributors for the new upper mid-band FR3 spectrum. To solve the current problem with industry standards bodies that are presently unable to take advantage of all the expensive and valuable measurements of contributors, here we propose a method for presenting and sharing simultaneously the statistical and site-specific nature of measured parameters, so that traditional channel parameters such as those given in \cite{Shakya2024gc, Shakya2024gc2, Shakya2024ojcoms} may be reproduced and augmented with additional data by any third party to create more accurate parameters. 
	
	To provide point-data that allows others to append the empirical results in the foundation of a larger pool of measurements, one merely needs to provide a site-specific representation of the environment (e.g., a map or a computerized model of the environment showing locations of the TX, RX and key obstructions and reflecting items) along with an agreed-upon format of a multidimensional table that specifies the LOS/NLOS scenario at each location, the TX-RX distance as drawn as a straight line through obstructions, and key statistics measured from the local power delay profile and angular profiles measured at each specified location, or point, within the environment. By implementing the point-data format (see Table \ref{tab:LSPs}), it becomes possible to use measured propagation data (e.g., channel parameters measured at a point in space, with such location in space denoted on a map or computerized representation of space) in order to characterize any extensive propagation measurement campaign in a standardized, site-specific manner for easy pooling of data. Such a multidimensional table (See Section \ref{PointData}, and Table \ref{tab:LSPs} specifically, which could be appended with contributions from many contributors) readily permits the computation of traditional CDFs and scatter plots by computing statistics down each column of a point-data table, but also offers a powerful way to create data for use in building accurate ray-tracing models for specific locations measured by many contributors, or to use pooled data sets for artificial intelligence and machine learning of both statistical and site-specific models from reported measurements. The following section demonstrates the proposed method, using extensive measurements in \cite{Shakya2024gc, Shakya2024ojcoms, Shakya2025wcnc, Ying2025tcom, Shakya2025icc}. 
	
	Before presenting the point-data representation, however, it should be noted that channel sounding systems of different bandwidths or different antenna patterns, as well as different thresholding/processing approaches, will have different degrees of resolution when providing point measurements \cite{ben2011millimeter, Ying2025tcom, Shakya2024TAP,Nie2013pimrc,ju2019icc}, so the minimum bandwidth should be specified in pooled data sets such that the lowest common denominator (e.g., the coarsest resolution, or the requirement of omni angular statistics having antenna patterns de-embedded) should be used by all contributors, or used as the baseline\cite{rappaport2012rws,rappaport2017investigation}. Nevertheless, basic channel parameters, such as bandwidth, antenna beamwidth, noise-threshold level, coarseness in point-data format, and time resolutions, as defined and used by 3GPP as defined in Section \ref{PointData}, could be provided on a best-effort basis, all in point-data format, and CDFs and scatter plots could readily be created and updated from the growing pool of data whenever additional data is supplied by a company or contributor in standardized point-data format. 
	
	\section{Point-Data for InH Measurements \cite{Shakya2024gc, Shakya2025wcnc, Shakya2024ojcoms}}
	\label{PointData}

	{\color{black}
		The data points in Table \ref{tab:LSPs} represent the point-data for the NYU WIRELESS InH measurement campaign \cite{Shakya2024gc, Shakya2024ojcoms, Shakya2025wcnc}, and summarize large-scale spatio-temporal statistics from the indoor office InH channel measurements at the NYU WIRELESS center, at the points in space shown in Fig. \ref{fig:measurement_maps} (a). \textcolor{black}{Similar point-data tables for UMi and InF are published in \cite{Ying2025tcom}.} The measurement campaign spanned 20 TX-RX locations with seven in LOS and 13 in NLOS, as noted in the first five columns of Table \ref{tab:LSPs}. Each column presents specific calculated statistics for large-scale parameters at the particular points in space denoted by columns 2 and 3 that describe the wireless propagation behavior and may be used for understanding particular site-specific channel parameters at specified TX, and RX points in space, and also, over all measured locations (i.e. across the entire Table \ref{tab:LSPs}), for generating CDFs and curve fitting for model parameterization. 
		The key parameters shown in Table \ref{tab:LSPs}, that are vital for use in 3GPP and NGA and other standard body proceedings, are described as follows:
		
		
		\renewcommand{\arraystretch}{1.07}
		\begin{table*}[!t]
			\centering
			\color{black}
			\caption{Point-data table for site-specific (InH) large scale spatio-temporal statistics with a map in Figure \ref{fig:measurement_maps} (a) \cite{Shakya2025wcnc,Shakya2024ojcoms} }
			\vspace{-0.2cm}
			\begin{tabular}{p{0.6 cm}p{0.5 cm}p{0.5 cm}p{0.69 cm}p{0.6 cm}p{0.69 cm}p{0.69 cm}p{0.69 cm}p{0.69 cm}p{0.69 cm}p{0.69 cm}p{0.69 cm}p{0.69 cm}p{0.6 cm}p{0.6 cm}p{0.6 cm}p{0.6 cm}}
				\hline
				\multicolumn{1}{p{0.6 cm}}{\textbf{Freq.}} & \textbf{TX} & \textbf{RX} & \textbf{Loc.} & \multicolumn{1}{p{0.6 cm}}{\textbf{TR Sep.}} & \multicolumn{1}{p{0.69 cm}}{\textbf{Omni Abs. PL (V-V)}} & \multicolumn{1}{p{0.69 cm}}{\textbf{Omni Abs. PL (V-H)}} & \multicolumn{1}{p{0.69 cm}}{\textbf{Mean Dir. DS}} & \multicolumn{1}{p{0.69 cm}}{\textbf{Omni DS}} & \multicolumn{1}{p{0.69 cm}}{\textbf{Mean Lobe ASA}} & \multicolumn{1}{p{0.69 cm}}{\textbf{Omni ASA}} & \multicolumn{1}{p{0.69 cm}}{\textbf{Mean Lobe ASD}} & \multicolumn{1}{p{0.69 cm}}{\textbf{Omni ASD}} & \multicolumn{1}{p{0.6 cm}}{\textbf{Mean Lobe ZSA}} & \multicolumn{1}{p{0.6 cm}}{\textbf{Omni ZSA}} & \multicolumn{1}{p{0.6 cm}}{\textbf{Mean Lobe ZSD}} & \multicolumn{1}{p{0.6 cm}}{\textbf{Omni ZSD}} \\
				\hline
				\text{[GHz]}& & & &[m]&[dB]&[dB]&[ns]&[ns]&[$^\circ$]&[$^\circ$]&[$^\circ$]&[$^\circ$]&[$^\circ$]&[$^\circ$]&[$^\circ$]&[$^\circ$] \\ 
				\hline
				\multirow{20}{*}{\textbf{6.75}} & \multirow{4}{*}{TX1} & RX1   & LOS   & 24.6  & 68.2  & 89.3 & 24.6  & 21.4  & 13.0  & 13.0  & 10.4  & 38.8  & 1.7   & 12.8  & 4.1   & 11.1 \\
				&       & RX2   & NLOS  & 27    & 88.6  & 102.5 & 23.7  & 77.4  & 47.8  & 47.8  & 12.9  & 51.3  & 6.8   & 8.3   & 3.0   & 10.5 \\
				&       & RX3   & NLOS  & 37.6  & 89.9  & 101.5 & 80.0  & 62.6  & 48.8  & 48.8  & 13.7  & 21.6  & 14.5  & 17.3  & 4.7   & 11.9 \\
				&       & RX4   & LOS   & 51    & 79.2  & 94.2  & 16.2  & 47.5  & 7.7   & 63.3  & 12.4  & 100.7 & 8.6   & 19.1  & 7.3   & 12.2 \\
				\cline{2-17}
				& \multirow{6}{*}{TX2} & RX1   & LOS   & 64    & 73    & 86.7  & 27.9  & 100.0 & 8.7   & 66.0  & 7.4   & 68.9  & 7.3   & 16.4  & 2.9   & 11.4 \\
				&       & RX2   & LOS   & 37.2  & 70.7  & 80.8  & 27.3  & 69.6  & 13.6  & 13.6  & 23.0  & 87.5  & 3.1   & 13.6  & 2.3   & 10.1 \\
				&       & RX3   & NLOS  & 67.3  & 103   & 114.2 & 33.3  & 47.0  & 11.4  & 56.5  & 6.4   & 62.8  & 7.3   & 16.2  & 6.4   & 15.8 \\
				&       & RX4   & NLOS  & 32.2  & 98.3  & 114.9 & 35.0  & 50.5  & 26.0  & 69.5  & 24.0  & 63.2  & 2.5   & 13.4  & 3.8   & 11.8 \\
				&       & RX5   & NLOS  & 49    & 111.2 & --  & 50.6  & 67.0  & 60.0  & 60.0  & 14.6  & 51.4  & 1.8   & 1.8   & 5.6   & 5.5 \\
				&       & RX6   & LOS   & 97    & 72.2  & 90.7  & 42.5  & 58.0  & 6.2   & 78.8  & 6.0   & 74.9  & 6.2   & 20.9  & 6.0   & 15.4 \\
				\cline{2-17}
				& \multirow{6}{*}{TX3} & RX1   & NLOS  & 12.3  & 72.6  & 85.1  & 7.3   & 27.7  & 14.4  & 58.2  & 19.5  & 80.8  & 8.4   & 18.0  & 6.2   & 12.0 \\
				&       & RX2   & NLOS  & 22.3  & 94.9  & 108.4 & 24.5  & 62.2  & 9.6   & 64.5  & 4.4   & 35.4  & 4.4   & 18.0  & 3.3   & 11.9 \\
				&       & RX3   & NLOS  & 65.6  & 88.8  & 100.8 & 18.9  & 32.9  & 41.7  & 78.6  & 9.5   & 65.3  & 10.5  & 11.8  & 7.3   & 13.3 \\
				&       & RX4   & NLOS  & 30    & 90.1  & 99  & 26.2  & 37.4  & 56.6  & 56.6  & 15.4  & 42.8  & 1.8   & 7.8   & 8.0   & 14.1 \\
				&       & RX5   & LOS   & 19.3  & 64.8  & 83.4  & 13.3  & 9.1   & 10.7  & 10.7  & 4.7   & 57.3  & 4.4   & 15.3  & 5.4   & 10.8 \\
				&       & RX6   & NLOS  & 49.5  & 105   & --  & 14.6  & 18.8  & 10.8  & 10.8  & 12.5  & 19.8  & 4.3   & 4.3   & 14.3  & 14.1 \\
				\cline{2-17}
				& \multirow{4}{*}{TX4} & RX1   & LOS   & 11.4  & 59    & 61.5  & 6.3   & 20.8  & 17.8  & 95.6  & 11.3  & 75.5  & 11.5  & 15.6  & 4.5   & 12.3 \\
				&       & RX2   & NLOS  & 17.1  & 68    & 83.7  & 20.6  & 41.9  & 59.4  & 59.4  & 24.8  & 84.5  & 1.8   & 9.0   & 7.0   & 12.6 \\
				&       & RX3   & NLOS  & 87    & 94.2  & --  & 6.3   & 23.4  & 17.8  & 73.1  & 6.6   & 33.8  & 4.9   & 18.0  & 6.6   & 14.4 \\
				&       & RX4   & NLOS  & 23.5  & 73.2  & 90.9  & 28.5  & 88.3  & 21.3  & 98.3  & 11.2  & 55.9  & 2.7   & 13.5  & 2.9   & 9.2 \\
				\hline
				\hline
				\multirow{20}{*}{\textbf{16.95}} & \multirow{4}{*}{TX1} & RX1   & LOS   & 24.6  & 73.8  & 97.5  & 14.0  & 15.6  & 8.6   & 8.6   & 5.9   & 73.6  & 6.3   & 8.5   & 4.5   & 8.3 \\
				&       & RX2   & NLOS  & 27    & 98.9  & 120.9 & 8.7   & 40.4  & 9.5   & 75.9  & 10.1  & 43.3  & 9.3   & 9.7   & 5.4   & 8.0 \\
				&       & RX3   & NLOS  & 37.6  & 100.9 & 110.3 & 18.4  & 51.3  & 7.8   & 55.2  & 8.7   & 16.4  & 4.4   & 10.7  & 8.5   & 8.4 \\
				&       & RX4   & LOS   & 51    & 83.8  & 102.2 & 18.4  & 10.0  & 7.1   & 7.1   & 6.8   & 90.8  & 3.7   & 9.5   & 4.9   & 8.7 \\
				\cline{2-17}
				& \multirow{6}{*}{TX2} & RX1   & LOS   & 64    & 82.6  & 102 & 22.2  & 21.3  & 6.0   & 72.2  & 4.2   & 64.1  & 5.4   & 12.8  & 4.5   & 9.3 \\
				&       & RX2   & LOS   & 37.2  & 79.3  & 100  & 22.6  & 27.3  & 7.2   & 7.2   & 4.7   & 62.2  & 9.1   & 11.1  & 4.4   & 7.9 \\
				&       & RX3   & NLOS  & 67.3  & 109.7 & 130.2 & 37.7  & 46.6  & 19.7  & 19.7  & 6.0   & 82.4  & 2.2   & 6.5   & 6.8   & 10.1 \\
				&       & RX4   & NLOS  & 32.2  & 107.9 & 142.3 & 13.2  & 45.1  & 7.1   & 62.8  & 14.3  & 108.4 & 3.6   & 11.3  & 3.1   & 6.6 \\
				&       & RX5   & NLOS  & 49    & 124.4 & --      & 4.9   & 16.5  & 5.2   & 64.9  & 12.7  & 12.7  & 5.2   & 7.5   & 11.8  & 11.8 \\
				&       & RX6   & LOS   & 97    & 80.3  & 110.8 & 41.9  & 60.2  & 7.3   & 7.3   & 4.1   & 63.6  & 4.1   & 9.3   & 4.4   & 8.9 \\
				\cline{2-17}
				& \multirow{6}{*}{TX3} & RX1   & NLOS  & 12.3  & 87.3  & 108.9 & 5.2   & 23.8  & 5.0   & 74.6  & 8.6   & 95.0  & 5.6   & 9.0   & 4.2   & 7.5 \\
				&       & RX2   & NLOS  & 22.3  & 111.3 & 132.0 & 25.6  & 65.7  & 11.8  & 11.8  & 7.8   & 57.8  & 2.0   & 7.5   & 3.6   & 3.9 \\
				&       & RX3   & NLOS  & 65.6  & 100.5 & 117.7 & 19.3  & 31.3  & 26.4  & 81.3  & 6.6   & 53.1  & 8.7   & 9.2   & 6.3   & 8.4 \\
				&       & RX4   & NLOS  & 30    & 96.8  & 109 & 14.4  & 51.6  & 20.8  & 58.9  & 10.1  & 12.0  & 8.0   & 8.7   & 4.6   & 8.2 \\
				&       & RX5   & LOS   & 19.3  & 73.2  & 91.2  & 12.1  & 9.1   & 6.9   & 6.9   & 7.2   & 39.0  & 7.9   & 10.6  & 4.1   & 7.4 \\
				&       & RX6   & NLOS  & 49.5  & 119.7 & 145.4 & 11.5  & 17.8  & 9.3   & 9.3   & 10.1  & 69.9  & 8.6   & 11.0  & 7.0   & 10.2 \\
				\cline{2-17}
				& \multirow{4}{*}{TX4} & RX1   & LOS   & 11.4  & 67.2  & 92.7  & 2.1   & 11.4  & 6.4   & 63.6  & 12.7  & 54.0  & 7.0   & 9.9   & 7.5   & 6.5 \\
				&       & RX2   & NLOS  & 17.1  & 93.3  & 114.6 & 15.1  & 52.1  & 14.1  & 109.9 & 14.8  & 73.5  & 3.0   & 7.3   & 3.8   & 6.6 \\
				&       & RX3   & NLOS  & 87    & 110.7 & 137.2 & 9.7   & 9.4   & 12.7  & 53.1  & 13.9  & 81.2  & 8.4   & 15.0  & 7.8   & 12.3 \\
				&       & RX4   & NLOS  & 23.5  & 91.9  & 110.1 & 17.8  & 76.9  & 14.2  & 95.1  & 20.3  & 52.1  & 2.5   & 7.5   & 2.6   & 4.6 \\
				\hline
			\end{tabular}%
			\label{tab:LSPs}%
			\vspace{-15pt}
		\end{table*}%
		\vspace{-5 pt}
		
		\medskip
		\renewcommand{\arraystretch}{1}
		\begin{itemize}
			\item \textbf{Freq.}: Center frequency in GHz.
			\item \textbf{TX}: TX location in the InH environment map, shown in Fig. \ref{fig:measurement_maps}(a).
			\item \textbf{RX}: RX location in the InH environment map corresponding to each TX location, shown in Fig. \ref{fig:measurement_maps}(a).
			\item \textbf{Loc.}: location type as being `LOS' or `NLOS'.
			\item \textbf{TR Sep.}: straight-line (Euclidean) distance between the TX and RX locations, measured in meters on the map. 
			\item \textbf{Omni Abs. PL (V-V)}: absolute omnidirectional path loss measured at a TX-RX location with antenna gains removed, expressed in dB, when using vertically polarized antennas at both TX and RX \cite{Shakya2024ojcoms}.
			\item \textbf{Omni Abs. PL (V-H)}: the path loss measured with cross-polarized antennas at a TX-RX location, where the vertical polarization at the transmitter and horizontal polarization at the receiver are considered. Antenna gains are removed from consideration, and the result is expressed in dB.
			\cite{Shakya2024ojcoms}.
			\item \textbf{Mean Dir. DS}: mean of the directional delay spreads from the directional PDPs captured at the RX for each TX-RX location pair, expressed in nanoseconds \cite{Shakya2024ojcoms}.
			\item \textbf{Omni DS}: omnidirectional delay spread at the RX for each TX-RX location pair, expressed in nanoseconds \cite{Shakya2024ojcoms}.
			\item \textcolor{black}{\textbf{Mean Lobe ASA}: mean of the logarithm of lobe azimuth spread in the multiple spatial lobes (SL) in the RX AOA power azimuth spectrum (PAS), with antenna pattern de-embedded and expressed in degrees, at each TX-RX location pair \cite{Shakya2025wcnc}.}\vspace{-5 pt}
			\begin{equation}
				\text{Mean Lobe ASA }= 10^{(1/L)\times\sum_l \log_{10}(\text{Lobe ASA}_l)}
				\vspace{-5 pt}
			\end{equation} 
			\textcolor{black}{where, Lobe ASA$_l$ denotes AS of $l^{th}$ SL among $L$ lobes in the RX AOA PAS.}
			
			\item \textbf{Omni ASA}: azimuth spread of arrival in the RX AOA PAS at each TX-RX location pair, expressed in degrees \cite{Shakya2025wcnc}. In Table \ref{tab:LSPs}, the AS is computed using the 3GPP method~\cite{3GPPTR38901}; a comprehensive comparison of the 3GPP method with the traditional Fleury approach is presented in \cite{Ying2025tcom}.
			\item \textcolor{black}{\textbf{Mean Lobe ASD}: mean of the logarithm of lobe azimuth spreads in the multiple SLs in the TX AOD PAS, with antenna pattern de-embedded and expressed in degrees, at each TX-RX location pair \cite{Shakya2025wcnc}.}
			\begin{equation}
				\text{Mean Lobe ASD }= 10^{(1/L)\times\sum_l \log_{10}(\text{Lobe ASD}_l)},
			\end{equation}
			\textcolor{black}{where, Lobe ASD$_l$ denotes AS of $l^{th}$ TX AOD PAS SL.}
			
			\item \textbf{Omni ASD}: azimuth spread of departure for the TX AOD PAS at each TX-RX location pair, expressed in degrees \cite{Shakya2025wcnc}.
			\item \textcolor{black}{\textbf{Mean Lobe ZSA}: mean of the log of lobe zenith spreads (ZS) in the multiple SLs in the RX AOA PAS at each TX-RX location pair, with antenna pattern de-embedded and expressed in degrees \cite{Shakya2025wcnc}.}\vspace{-3 pt}
			\begin{equation}
				\text{Mean Lobe ZSA }= 10^{(1/L)\times\sum_l \log_{10}(\text{Lobe ZSA}_l)},
			\end{equation}
			\textcolor{black}{where, Lobe ZSA$_l$ denotes ZS of $l^{th}$ RX AOA PAS SL.}
			
			\item \textbf{Omni ZSA}: zenith spread of arrival for the RX AOA PAS at each TX-RX location pair, expressed in degrees \cite{Shakya2025wcnc}.
			\item \textcolor{black}{\textbf{Mean Lobe ZSD}: mean of the logarithm of lobe zenith spreads in the multiple SLs in the TX AOD PAS at each TX-RX location pair, with antenna pattern de-embedded and expressed in degrees \cite{Shakya2025wcnc}.}\vspace{-3 pt}
			\begin{equation}
				\text{Mean Lobe ZSD }= 10^{(1/L)\times\sum_l \log_{10}(\text{Lobe ZSD}_l)},
			\end{equation}
			\textcolor{black}{where, Lobe ZSD$_l$ denotes ZS of $l^{th}$ TX AOD PAS SL.}
			
			\item \textbf{Omni ZSD}: zenith spread of departure for TX AOD PAS at each TX-RX location pair, expressed in degrees \cite{Shakya2025wcnc}.
		\end{itemize}

		The channel parameters and the point-specific values given in Table \ref{tab:LSPs}, when combined with the InH map given in Fig. \ref{fig:measurement_maps} (a), offer a new way for researchers to supply propagation measurement data in a way that enables point-level understanding, which we call the ``point-data'' format. Different contributors could simply append data by adding rows to Table \ref{tab:LSPs}, where the statistics for the key channel parameters in each column can be updated to yield improved statistical models from the ever-growing dataset. 
		\color{black}{It is important to note that the conventional method of creating CDFs or scatter plots may easily be performed using Table \ref{tab:LSPs}, using only the columns of Table \ref{tab:LSPs}, as shown in \cite{Shakya2024ojcoms,Ying2025tcom}.} 
	}

	\section{Conclusion}
	Extensive work is ongoing to determine statistical channel models for the FR3 mid-band spectrum. In this paper, we highlighted the activity to measure and model the FR3 bands and presented a novel approach to allow the presentation of propagation data in a point-data format with site-specific information. If industry and academia followed such a convention, it would become extremely easy for standard bodies or other organizations to pool data sets from many sources for comparison, combination, assessment, and learning. The multidimensional point-data format support both traditional statistical analysis using CDFs and scatter plots while simultaneously enabling the development of accurate ray-tracing models and data-driven approaches using AI/ML techniques. Further, such a standard approach for data presentation could drive other standard-setting behaviors, such as bounding the time, frequency, and angular/spatial resolution of particular measurement campaigns for inclusion in data sharing, data pooling, and data aggregation.

	\section*{Acknowledgment}
	The authors thank Hitesh Poddar and others at Sharp Laboratories of America for discussions that led to the point-data format for 3GPP and other standard bodies.
	
	\bibliographystyle{IEEEtran}
	\bibliography{references}
	
	\end{document}